# Single-ion phonon laser in the quantum regime[*]

DONG Yuanzhang[1,2], HE Siwen[3], DENG Zhijiao[3], LI Peidong[1,2], CHEN Liang[1], FENG Mang[1]

1.Innovation Academy of Precision Measurement Science and Technology, Chinese Academy of Sciences, Wuhan 430071, China

2.University of the Chinese Academy of Sciences, Beijing 100049, China

3.College of Science, National University of Defense Technology, Changsha 410073, China

**Abstract**

The quantum phonon laser state is a vibrational state generated by phonon coherent amplification technology based on the principles of quantum mechanics. Its core feature is to achieve coherent excitation and manipulation of phonon quantum states through precise control of phonon dynamics. This technology has broken through the classical physical limits of the traditional phonon laser state, providing a brand-new research method for quantum information technology. Previous research on quantum phonon laser states mainly focused on quantum van der Pol oscillators. Quantum van der Pol oscillators, as typical representatives of nonlinear quantum systems, have demonstrated significant theoretical value and broad application prospects in trapped-ion systems in recent years. These research breakthroughs not only successfully expand the research scope of traditional nonlinear dynamics to the quantum domain, but more importantly, provide a brand-new experimental platform and theoretical framework for exploring quantum nonlinear phenomena.

Although the realization of quantum phonon laser state has been verified in two-ion systems, its practical application still faces significant challenges. The present paper explores how a single trapped ion generates quantum phonon laser states based on the three-level model. By numerically solving the quantum master equation, the steady-state characteristics of the phonon laser state are systematically analyzed, with a focus on the quantum statistical behavior of the system, including the evolution of the Wigner quasi-probability distribution function and the second-order correlation function. This paper also presents a

---

specific experimental scheme, which is based on a single trapped $^{40}Ca^+$ ion and uses a bichromatic light field composed of a blue-sideband and a red-sideband lasers to generate quantum phonon laser states. By introducing the characteristic function of motional quantum states, the precise quantum state tomography of phonon laser states is achieved, In addition, there is a two-level model discussing the threshold effect of phonon lasers. However, it is found that the three-level model constructed in the present paper has significantly different phonon laser thresholds compared with the two-level model, and the three-level model can describe more accurately the physical mechanisms of quantum phonon laser states.

## 1. Introduction

Phonon laser state in quantum region is a kind of vibrational state generated by phonon coherent amplification technology based on the principle of quantum mechanics. Its core feature is to realize the coherent excitation and manipulation of phonon quantum state through precise control of phonon dynamics. This technology breaks through the classical physical limit of the traditional phonon laser state and provides a new research method for quantum information technology, which is mainly manifested in three aspects. Firstly, in the aspect of quantum information processing, it realizes the precise control of the quantum states of phonon-photon coupled systems; secondly, in the field of precision measurement, its quantum enhancement feature significantly improves the sensitivity of nanoscale force sensing; finally, it provides an experimental platform for novel quantum states such as quantum coherence, superfluid state and topological phonon state. At the level of theoretical description, the van der Pol model can qualitatively explain the steady-state behavior and threshold characteristics of phonon laser, while for the quantum effects of the system, such as quantum noise and energy level discretization, Lindblad master equation or quantum fluctuation theory should be further introduced for more detailed modeling and analysis. The single-ion phonon laser state in trapped-ion system is mainly based on the van der Pol oscillator mechanism, which has shown unique advantages in the field of weak signal detection, such as high sensitivity and low noise. This highly controllable quantum acoustic field also provides an ideal tool for studying the dynamic behavior of complex quantum systems. It is worth noting that this interdisciplinary research field integrates the frontier achievements of quantum physics, optical engineering, phononics and advanced materials science, which not only promotes the development of basic research fields, but also opens up new ways for the development of application technologies such as quantum computing, quantum communication and quantum sensing.

Previous studies on phonon laser states in the quantum region were mainly carried out around the quantum van der Pol oscillator. As a typical representative of nonlinear quantum systems, quantum van der Pol oscillator has shown important theoretical value and broad application prospects in trapped-ion systems in recent years. These research breakthroughs not only successfully extend the research scope of traditional nonlinear dynamics to the quantum field, but also provide a new experimental platform and theoretical framework for exploring quantum nonlinear phenomena. From the theoretical point of view, a series of breakthroughs have been made in this field: the first observation of quantum limit cycle reveals the self-sustained oscillation characteristics in quantum systems, while the discovery of nonlinear quantum dissipation mechanism provides a new perspective for understanding the energy dissipation process of quantum systems. These pioneering studies have not only deepened the understanding of the essential characteristics of quantum nonlinear systems, but also laid an important foundation for the development of frontier technologies such as quantum information processing and quantum simulation. In 2013, Lee and Sadeghpour systematically studied the synchronization phenomenon of quantum van der Pol oscillator through theoretical modeling. Their study demonstrated that under the same parameters, the quantum system showed more significant phase locking robustness than the classical system. Furthermore, they proposed a specific experimental scheme based on a trapped-ion system, along with recommendations for parameter optimization. In 2019, Dutta and Cooper revealed the unique quantum characteristics of quantum van der Pol oscillator through numerical simulation and analytical calculation. The latest research in 2023 has confirmed that topological effects can significantly enhance quantum synchronization. In the topological lattice of quantum van der Pol oscillators, boundary synchronization not only exists at the classical mean-field level, but also shows clear experimental observable features in the quantum system.

With the first observation in a trapped-ion system, "phonon laser" has been successfully realized in a variety of physical systems, covering many research platforms from atomic systems to nanomechanical systems. It is particularly striking that the amplitude of oscillations observed in atomic systems can be of the order of tens of micrometers, which corresponds to phonon numbers of the order of about $10^4$. In recent years, phonon laser states have made a series of important breakthroughs in the fields of quantum control and nonlinear dynamics. In 2020, Flühmann and Home innovatively mapped the expectation value of the real or imaginary part of the displacement operator to the internal state of the ion through a specific internal state selection technique, thereby enabling the direct measurement of the characteristic function of the vibrational quantum state. Furthermore, they established a new paradigm for characterizing vibrational quantum states by reconstructing the complete Wigner function through two-dimensional Fourier transform. In 2023, Behrle et al. made a breakthrough in the field of quantum control by using a mixed-species dual-ion system to construct a double dissipation channel, thus realizing a stable phonon laser state in the quantum region for the first time. This work not only experimentally verified the basic principle of quantum phonon laser, but also opened up a new direction for quantum acoustic field research. In 2024, He Siwen studied the quantum synchronization characteristics of single-ion phonon laser state by controlling the external driving field, and observed significant oscillations in the evolution of entanglement dynamics, which provided a new theoretical basis for understanding the correlation mechanism between quantum synchronization and quantum entanglement.

These research achievements have not only advanced the development of the theory of quantum phonon dynamics but also opened new avenues for the engineering applications of quantum technologies.With the continuous innovation of experimental methods and theoretical tools, phonon laser states show more and more unique advantages in quantum computing, precision sensing and complex system simulation. The most striking feature of the quantum phonon laser state is that it breaks through the limitations of classical physics, and can achieve measurement sensitivity close to the quantum limit and excellent steady-state characteristics. These properties make it the core component of future high-precision measurement instruments. In quantum systems, the realization of phonon laser states has been demonstrated in two-ion systems, but its practical application still faces significant challenges. This scheme requires that the mixed-species dual-ion system must exactly share the vibrational center-of-mass mode, and the Coulomb interaction between ions may lead to the phenomenon of phonon mode hybridization. In order to overcome these problems, it is necessary to design a complex laser control scheme and improve the accuracy of spectrum control, which greatly increases the complexity of the experimental system. In contrast, the single-ion system has obvious advantages: its structure is simple, the phonon mode purity is high, and the manipulation of a single quantum state is more convenient. In addition, the single-ion scheme not only simplifies the laser configuration requirements, but also significantly improves the reliability and experimental repeatability of the system. However, in determining the threshold of phonon laser generation, the analysis process for the single-ion phonon laser needs to consider the quantum process between two excited states, while two-ion phonon laser does not need to consider, so the determination of single-ion phonon laser threshold is relatively more complex. However, under the condition of external driving, the internal and external degrees of freedom of a single-ion system will form steady-state entanglement, which enables the information of the external state to be obtained synchronously by measuring the internal state, providing a new idea for quantum state detection.

Although the two-level model of the single-ion scheme provides a concise theoretical framework for the quantum phonon laser state, its oversimplified energy level structure also has significant limitations. This model considers only the ground state and the excited state, which can not accurately describe the multi-level structure characteristics commonly found in real atoms or molecules, such as intermediate levels, metastable states and other important energy states. This simplification makes the model have essential limitations in describing multi-level quantum transition processes, especially in accurately characterizing key dynamic behaviors such as the diversity of relaxation paths and non-radiative transition channels, which leads to systematic deviations between theoretical predictions and experimental observations when explaining complex physical phenomena such as population inversion mechanism and laser dynamics. In contrast, the theoretical model of the single-ion scheme is extended to the three-level model, which significantly improves the physical reality and descriptive ability of the model by introducing additional energy level structures. Different from the two-level model which usually requires the adiabatic approximation, the three-level model relaxes these restrictive conditions and can reflect the characteristics of the actual quantum system more comprehensively. This improvement allows the three-level model to accurately describe the physical mechanism of more complex quantum phonon laser states. More importantly, the predictions of the three-level model are in better quantitative agreement with experimental observations, which makes it an effective theoretical tool for studying practical laser systems and quantum optical devices.

In this paper, a three-level theoretical model of the phonon laser state in the quantum regime of a single ion is constructed. By solving the quantum master equation numerically, the steady-state characteristics of the phonon laser state are systematically analyzed. The quantum statistical behavior of the system is emphatically studied, including the evolution of the Wigner quasi-probability distribution function and the second-order correlation function. Based on a trapped single $^{40}Ca^+$ ion, a bichromatic light field composed of blue-sideband laser and red-sideband laser is proposed to generate quantum phonon laser state. By introducing the characteristic function of the motional quantum state, the precise quantum state tomography of the phonon laser state is realized. In addition, although both the three-level model and the two-level model show obvious phonon laser threshold effect, a direct comparison under identical parameters is not feasible. Therefore, this paper shows the significant deviation between the three-level model and the two-level model by comparing their phonon laser generation thresholds, which indicates that the threshold conditions are model-dependent and cannot be simply interchanged.

## 2. Three-level model

In a trapped-ion system, the ion has two independent degrees of freedom: the electron spin state and the ion vibrational state. These two degrees of freedom can be effectively coupled by the laser field. The ion vibrates with frequency $\nu$ in the potential well shown by Fig 1(a), and the ion maintains the stable amplitude of the phonon laser state during the competitive process of heating and cooling. When the laser frequency is close to resonance with the transition frequency between the two levels of the ion, the system can be simplified to a three-level model consisting of $|g\rangle$, $|e_1\rangle$ and $|e_2\rangle$ as shown in Fig 1(b). A laser with a detuning of $\delta = -\nu$ is called a red-sideband laser; A laser detuned by $\delta = \nu$ is called a blue-sideband laser. Driven by the blue-sideband laser, the ion transits from the ground state $|g\rangle$ to the excited state $|e_1\rangle$, accompanied by the generation of a phonon; Subsequently, the ion returns from the $|e_1\rangle$ state to the $|g\rangle$ state at a decay rate $\gamma_h$ through an effective dissipation process, which realizes the blue-sideband heating of the ion. On the other hand, under the action of the red-sideband laser, the ion transits from the ground state $|g\rangle$ to the excited state $|e_2\rangle$ and absorbs a phonon at the same time; Then, the ions return from the $|e_2\rangle$ state to the $|g\rangle$ state through an effective dissipation process at a decay rate $\gamma_c$, thus realizing the red-sideband cooling of the ions.

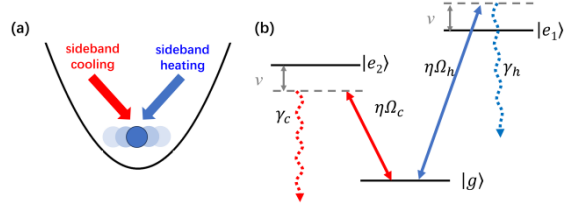

**Figure 1.** Three-level model for generating quantum phonon laser. (a) Schematic diagram of phonon laser generated by two-color light field. The ion vibrates in the potential well, where the blue-sideband heating and red-sideband cooling processes of the ion are represented by blue and red arrows, respectively, and the heating and cooling compete with each other to maintain a stable amplitude of the phonon laser. (b) The internal energy level structure of the ion.

When the amplitude of the ion motion is small, its motion can be approximated as a simple harmonic vibration, and the total Hamiltonian $H$ of the ion is composed of two parts, the free Hamiltonian $H_0$ and the laser-ion interaction $H_h$ and $H_c$, namely

$$\hat{H} = \hat{H}_0 + \hat{H}_h + \hat{H}_c, \tag{1}$$

$$\hat{H}_0 = \frac{1}{2}\omega_{0h}\hat{\sigma}_{zh} + \frac{1}{2}\omega_{0c}\hat{\sigma}_{zc} + v\hat{a}^\dagger\hat{a}, \tag{2}$$

$$\hat{H}_h = \Omega_h \hat{\sigma}_h^+ \{e^{-i\omega_{Lh}t}[1 + i\eta(\hat{a}^\dagger + \hat{a})]\} + \text{h.c.}, \tag{3}$$

$$\hat{H}_c = \Omega_c \hat{\sigma}_c^+ \{e^{-i\omega_{Lc}t}[1 + i\eta(\hat{a}^\dagger + \hat{a})]\} + \text{h.c.}, \tag{4}$$

where $\Omega_h$ and $\Omega_c$ are the Rabi frequencies of the interaction of the blue-sideband laser and the red-sideband laser with the ion, respectively; The $\eta$ is Lamb-Dicke coefficient; $\hat{a}(\hat{a}^\dagger)$ is the annihilation (creation) operator of the ion's vibrational mode; The $v$ is the vibration frequency of the ion in the potential well; $\omega_{0h}$ is the transition frequency between the ground state $|g\rangle$ and the excited state $|e1\rangle$; $\omega_{0c}$ is the transition frequency between the ground state $|g\rangle$ and the excited state $|e_2\rangle$; $\omega_{Lh}$ is the laser frequency of the blue-sideband laser; $\omega_{Lc}$ is the laser frequency of the red-sideband laser; $\hat{\sigma}_{zh} = |e_1\rangle\langle e_1| - |g\rangle\langle g|$, $\hat{\sigma}_{zc} = |e_2\rangle\langle e_2| - |g\rangle\langle g|$, $\hat{\sigma}_h = |g\rangle\langle e_1|$ describe the process of $|e_1\rangle \rightarrow |g\rangle$, $\hat{\sigma}_c = |g\rangle\langle e_2|$ describes the process of $|e_2\rangle \rightarrow |g\rangle$, while $\hat{\sigma}_h^+$ and $\hat{\sigma}_c^+$ are the Hermitian conjugate operators of $\hat{\sigma}_h$ and $\hat{\sigma}_c$, respectively.

The total Hamiltonian $H$ of the ion is transformed under the interaction representation, i.e.

$$\hat{H}_I = \eta\Omega_h(\hat{a}^\dagger\hat{\sigma}_h^+ + \hat{a}\hat{\sigma}_h) + \eta\Omega_c(\hat{a}\hat{\sigma}_c^+ + \hat{a}^\dagger\hat{\sigma}_c). \tag{5}$$

Described by the following Lindblad master equation:

$$\dot{\hat{\rho}} = -i[H_I, \hat{\rho}] + \gamma_h D[\sigma_h]\hat{\rho} + \gamma_c D[\sigma_c]\hat{\rho}, \tag{6}$$

$$D[\hat{\sigma}_h]\hat{\rho} = \hat{\sigma}_h\hat{\rho}\hat{\sigma}_h^+ - \frac{1}{2}(\hat{\sigma}_h^+\hat{\sigma}_h\hat{\rho} + \hat{\rho}\hat{\sigma}_h^+\hat{\sigma}_h), \tag{7}$$

$$D[\hat{\sigma}_c]\hat{\rho} = \hat{\sigma}_c\hat{\rho}\hat{\sigma}_c^+ - \frac{1}{2}(\hat{\sigma}_c^+\hat{\sigma}_c\hat{\rho} + \hat{\rho}\hat{\sigma}_c^+\hat{\sigma}_c), \tag{8}$$

where $\gamma_h$ and $\gamma_c$ are the effective decay rates corresponding to the excited states $|e_1\rangle$ and $|e_2\rangle$, respectively; and $\hat{\rho}$ is the density matrix.

## 3. Simulation result

### 3.1 Phase diagram of quantum phonon laser state, Wigner quasi-probability distribution and phonon number distribution

By solving the Lindblad master equation (6) of the quantum system, the quantum properties of the phonon laser state will be analyzed in detail. Choose the parameters as $\gamma_c = 100$ and $\gamma_h = 1$, and let $g_c = \eta\Omega_c$, $g_h = \eta\Omega_h$. Each set of coupling parameters $g_c$ and $g_h$ corresponds to a specific Hamiltonian $H_I$, which is numerically solved by substituting these parameters into (6) to obtain the average phonon number $\langle\hat{n}\rangle = \langle\hat{a}^\dagger\hat{a}\rangle$ under the corresponding parameter conditions.

By systematically changing the coupling parameters, the steady-state phase diagram shown in Fig. 2(a) can be obtained, which can clearly illustrates the stability characteristics of the system in different parameter regions.

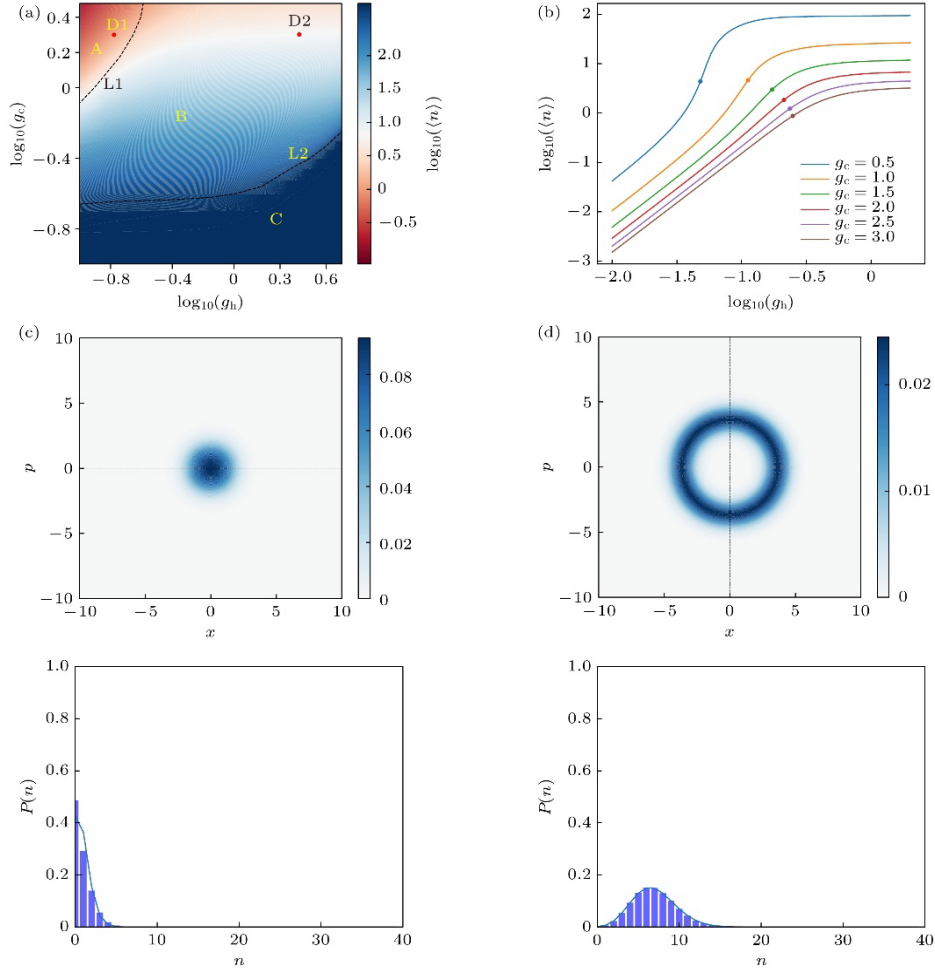

**Figure 2.** (a) Phase diagram, where the horizontal and vertical coordinates and the average number of phonons are denoted by logarithms with base 10. Number of truncation is 600. Region A is the thermal state region, region B is the phonon laser region, and region C is the diverging region. The dashed curve L1 is the boundary between the thermal state and the phonon laser, and the dashed curve L2 is the boundary between the converging and diverging regions; (b) the mean phonon number as a function of $g_h$. The average phonon number first goes up quickly, and then converges to a stable value; (c) D1(–0.8239, 0.301) steady-state Wigner quasi-probability distribution, with the parameters $g_h = 0.15, g_c = 2$; (d) D2(0.4771, 0.301) steady-state Wigner quasi-probability distribution, with the parameters $g_h = 3, g_c = 2$.

As shown in Fig. 2(a), the phase diagram can be clearly divided into three characteristic regions according to the threshold line L1 and the boundary line L2: region A, region B, and region C. Specifically, the A region is located to the left of the threshold line L1; The B region is distributed on the right side of the threshold line L1 and above the boundary line L2; Region C is below the dividing line L2. In the region A, the phonon number distribution shows the typical thermal distribution. In the B region, the ions enter the self-excited oscillation state, showing stable amplitude and random phase. The results in the C region is the region that has not been converged when the truncation number is 600, which includes the phonon laser state with phonon number greater than 600 and the unsteady state. It should be noted that the C region in the **Fig. 2(a)** is an unsteady result caused by insufficient number of iterations in the numerical calculation, so it can be ignored in the analysis of this paper.

In this paper, two methods are used to determine the threshold of the phonon laser state. The first method is shown in **Fig. 2(b)**, and the threshold line L1 between the thermal state and the phonon laser state is determined by analyzing the variation characteristics of the average phonon number $\langle n \rangle$ with the control parameters under different parameters. Specifically, when the system parameters cross the critical value, the average phonon number will have a significant jump, and the jump point is the transition threshold point. In order to accurately determine the critical point, the numerical differentiation method is used to take the partial derivative of the curve, and the extreme point of the derivative corresponds to the phase transition threshold of the system. By systematically scanning the parameter space and determining the phase transition threshold under each parameter combination, the complete threshold line L1 is finally constructed. On the other hand, the boundary line L2 of the system is determined by the long-term behavior of the phonon number evolution with time. In the stable region, the mean phonon number will rapidly converge to the steady-state value after a short relaxation process; In the unstable region, the average phonon number shows a continuous growth until the truncation condition of the numerical simulation is triggered. Based on the difference of the dynamic behavior, the system boundary L2 can be obtained when the Fock space truncation number is 600.

Another important method to judge the threshold of phonon laser state is to compare the difference between the Wigner quasi-probability distribution function and the phonon number distribution in the regions A and B. Wigner quasi-probability distribution function, as the core mathematical expression of quantum phase space theory, can be used to completely describe the phonon laser state in position-momentum phase space. Its general mathematical expression is

$$W(x,p) = \frac{1}{\pi\hbar} \int_{-\infty}^{\infty} \langle x+y| \hat{\rho} |x-y\rangle \, e^{2ipy/\hbar} dy, \tag{9}$$

where $x$ and $p$ denote the eigenvalues of the position operator $\hat{x} = (\hat{a} + \hat{a}^\dagger)/\sqrt{2}$ and the momentum operator $\hat{p} = (\hat{a} - \hat{a}^\dagger)/(\sqrt{2}i)$, respectively; $\hbar$ represents Planck's constant; $\hat{\rho}$ is the density matrix of the system. The analysis of phonon laser should focus on the phonon state of the ion. A mathematical correspondence exists between the Wigner quasi-probability distribution function of phonon state and the phonon number distribution function. The reduced density matrix $\hat{\rho}_{ph}$ of the external state can be obtained by partial trace of the density matrix of the system[33].

$$\hat{\rho}_{\text{ph}} = \text{Tr}_{\text{in}} \hat{\rho}. \tag{10}$$

The phonon number distribution can be reflected by the diagonal matrix elements:

$$P(n) = \langle n| \hat{\rho}_{\text{ph}} |n\rangle. \tag{11}$$

Based on this, the quantum regime with an average phonon number of less than 10 is selected to systematically analyze[30] and reveal the unique properties of phonon laser states in the quantum regime. Fig. 2(c) and Fig. 2(d) show the Wigner quasi-probability distribution of the typical characteristic points D1 and D2 on both sides of the threshold line L1, respectively. It is found that at point D1, the Wigner function exhibits a symmetric Gaussian distribution, which reflects the statistical characteristics of the system in thermal equilibrium; At D2, the Wigner function shows an obvious ring structure, forming a typical limit cycle distribution, accompanied by the Poisson statistical properties of the phonon number distribution, which clearly indicates that the system has entered the quantum coherent state region. This comparative analysis not only intuitively shows the difference of quantum statistical properties of the phonon state in the two parameter regions A and B, but also provides an important basis for understanding the phase transition process of the phonon laser.

In the region A to the left of the critical threshold, the coordinate and momentum of the ion tend to be stable after a long time evolution, so it appears as a fixed point in phase space. The corresponding Wigner quasi-probability distribution exhibits a highly symmetric circular spot structure, and its rotational symmetry indicates that the system is in thermal equilibrium. When the parameter crosses the critical threshold and enters the region B on the right, the system dynamics exhibits the typical characteristics of a nonlinear dissipative system, which forms a limit cycle of stable amplitude in the phase space. In this state, the Wigner quasiprobability distribution exhibits the phase space structure of the nonclassical state, and the phonon number distribution can be fitted by a Poisson distribution, that is,[33]

$$P(n) = \frac{\lambda^n}{n!} e^{-\lambda}, \tag{12}$$

where $\lambda$ represents the expectation value of the phonon, and the phonon number distribution changes from a Gaussian distribution to a Poisson distribution with discrete characteristics, which indicates that the system has entered the phonon laser state region. The accuracy of the threshold line L1 can be further determined by observing the difference between the Wigner quasi-probability distribution and the phonon number distribution in the two regions. Specifically, when the selected parameter exceeds the threshold line, the Wigner quasi-probability distribution of the phonon state exhibits a distinct ring structure, indicating the emergence of coherence.

Due to the additional level degrees of freedom introduced by the three-level system compared to the two-level system, the phase diagrams of the two systems cannot be directly compared. Although both models exhibit the typical phonon laser threshold effect[33] (which is manifested by the sudden change of the average phonon number near the critical point), the threshold conditions of both models show significant non-universal characteristics. Specifically, there is no simple scaling relationship between the threshold parameters of a three-level system and a two-level system. From the following discussion of the experimental scheme, it can be seen that the three-level system can better describe the energy level structure of $^{40}Ca^+$ ions as shown in **Fig. 4(a)**, so it can more accurately reflect the dynamic behavior of the actual physical system. In contrast, the two-level approximation can not accurately describe the dynamics in the experimental system because of the insufficient number of energy levels considered.

## 3.2 Second-order correlation function of quantum phonon laser states

The second-order correlation function of a phonon laser is a key metric for distinguishing between classical and quantum phonon behavior, as it differentiates a classical phonon field (e.g., thermal phonons) from a quantum phonon field (e.g., single-phonon sources or entangled phonon pairs). It is defined as[33]:

$$g^{(2)}(\tau) = \frac{\langle \hat{a}^\dagger(t)\hat{a}^\dagger(t+\tau)\hat{a}(t+\tau)\hat{a}(t) \rangle}{\langle \hat{a}^\dagger(t)\hat{a}(t) \rangle \langle \hat{a}^\dagger(t+\tau)\hat{a}(t+\tau) \rangle}, \tag{13}$$

where $\tau$ represents the time delay. This expression describes the intensity coherence of a single mode field at a fixed position in space at different times of $t$ and $t+\tau$. In the theoretical calculation of the second-order correlation function, the time delay $\tau$ is normalized to some characteristic time scale, in which case the $\tau$ is dimensionless[37]. In **Fig. 3(a)**, the parameter of the black line corresponds to the D1 point in the A region of **Fig. 2(a)**, and the initial value is $g^{(2)}(\tau) > 1$, indicating that the thermal phonon is dominant in the system. At this time, the statistical distribution of particles is more discrete than the Poisson distribution. Experimentally, this phenomenon manifests as anti-bunching characteristics in the temporal dimension of phonon detection events, meaning that phonons are more likely to be detected separately than to arrive simultaneously. With the increase of the time delay $\tau$, the second-order correlation function $g^{(2)}(\tau)$ shows an exponential decay trend and gradually approaches 1, and its decay time is related to the phonon intrinsic lifetime. The parameters of the red line correspond to the D2 point in the B region of **Fig. 2(a)**, and its $g^{(2)}(\tau)$ value is basically kept around 1, which shows typical coherence characteristics.

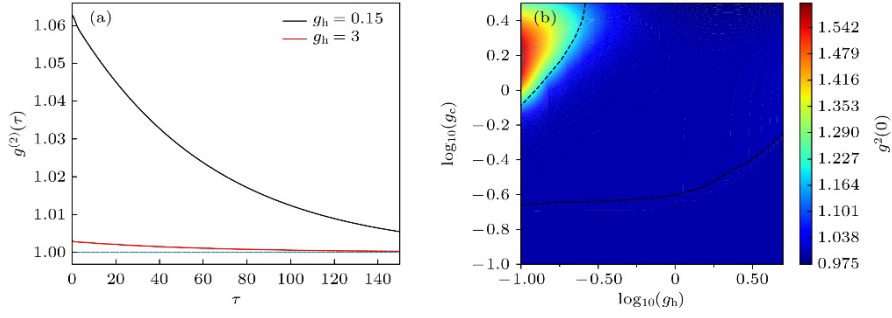

**Figure 3.** (a) Second-order correlation function $g^{(2)}(\tau)$ above and below the phonon laser threshold, where the parameter is $g_c=2$; (b) second-order correlation function $g^{(2)}(0)$ with zero time delay.

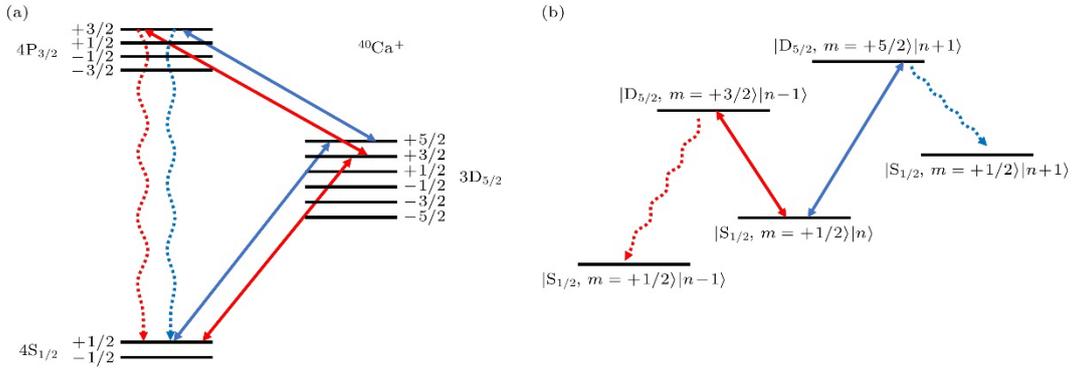

**Figure 4.** (a) Internal energy level scheme. The blue and red lines represent blue-sideband heating and red-sideband cooling, respectively. (b) Schematic diagram of the blue-sideband heating process and the red-sideband cooling process.

In the case of zero time delay ($\tau=0$), the $g^{(2)}(\tau)$ is reduced to $g^{(2)}(0)$ [33]:

$$g^{(2)}(0) = \frac{\langle \hat{n}^2 \rangle - \langle \hat{n} \rangle}{\langle \hat{n} \rangle^2}, \qquad (14)$$

where $\langle \hat{n} \rangle$ is the mean phonon number. As shown in Fig. 3(b), in region A, the $g^{(2)}(0) > 1$ indicates that there is bunching effect in the phonon system, which is manifested by the non-uniform clustering distribution of phonons in time and space dimensions, and its statistical characteristics deviate from Poisson distribution, showing obvious clustering fluctuation characteristics; In the B region, $g^{(2)}(0) \approx 1$, the system shows coherent properties, and the phonon number distribution is characterized by Poisson distribution.

## 4. Experimental scheme

### 4.1 Generation of phonon laser states in the quantum regime.

The energy levels of $^{40}Ca^+$ ions selected in the experiment are shown in Fig. 4(a). The study mainly focuses on the fine structure splitting states corresponding to the three energy levels of $4S_{1/2}$, $4P_{3/2}$ and $3D_{5/2}$, namely, $|4S_{1/2}, m = +1/2\rangle$ of $4S_{1/2}$, $|3D_{5/2}, m = +5/2\rangle$ and $|3D_{5/2}, m = +3/2\rangle$ of $3D_{5/2}$, and $|4P_{3/2}, m = +3/2\rangle$ of $4p_{3/2}$. The $|4P_{3/2}, m = +3/2\rangle$ of $4P_{3/2}$ is the auxiliary state of the blue-sideband heating process and the red-sideband cooling process. The specific experimental process is as follows.

Firstly, the motional degree of freedom of the $^{40}Ca^+$ ion is cooled to the ground state by sideband cooling, so that the decoherence effect caused by the thermal motion of the ion is effectively suppressed, and the foundation is laid for the subsequent quantum state manipulation. Driven by the blue-sideband of 729nm laser, the ion first transits from $|4S_{1/2}, m = +1/2\rangle \otimes |n\rangle$ state and generates a phonon to $|3D_{\frac{5}{2}}, m = +\frac{5}{2}\rangle \otimes |n + 1\rangle$ state; Then the ions are transferred to $|4P_{3/2}, m = +3/2\rangle \otimes |n + 1\rangle$ state by the 854nm laser, and finally returned to $|4S_{1/2}, m = +1/2\rangle \otimes |n + 1\rangle$ state through spontaneous emission, which realizes the blue-sideband heating mechanism of the ions.

On the contrary, driven by the red-sideband of the 729nm laser, the ion shows a different dynamic behavior from $|4S_{1/2}, m = +1/2\rangle \otimes |n\rangle$ state and annihilates a phonon to $|3D_{\frac{5}{2}}, m = +\frac{3}{2}\rangle \otimes |n - 1\rangle$ state; Then it is excited to $|4P_{3/2}, m=+3/2\rangle \otimes |n-1\rangle$ state by the 854nm laser, and finally spontaneously radiates back to $|4S_{1/2}, m=+1/2\rangle \otimes |n-1\rangle$ state, which constitutes the red-sideband cooling mechanism.

The above red-sideband cool and blue-sideband heating processes mainly involve three states of $|4S_{1/2}, m=+1/2\rangle$, $|3D_{5/2}, m=+5/2\rangle$ and $|3D_{5/2}, m=+3/2\rangle$, which respectively correspond to three states of $|g\rangle$, $|e_1\rangle$ and $|e_2\rangle$ as shown in Fig. 1(b); $|n\rangle$, $|n-1\rangle$ and $|n+1\rangle$ represent phonon states, the blue line represents the blue-sideband heating process and the red line represents the red-sideband cooling process, which can be combined to obtain a schematic diagram of Fig. 4(b). During the generation of the phonon laser state, the two competing mechanisms of blue-sideband heating and red-sideband cooling reach a dynamic balance, thus effectively maintaining the steady state of the ion's vibrational mode. This balance mechanism enables the system to maintain stable dynamic characteristics during the quantum manipulation process.

## 4.2 Measurement of phonon laser states in the quantum regime

The phonon laser state in the quantum regime can be characterized by measuring the characteristic function and obtaining the Wigner function through two-dimensional Fourier transform. The specific process is as follows.

First we prepare the ion's vibrational state to $|\psi\rangle$ and the electron spin state to $|4S_{1/2}, m=+1/2\rangle$, then the carrier transition $\hat{R}(\theta, \phi = 0)$ is applied on the electron spin state and the displacement operator $\hat{D}(\alpha)$ is applied on the ion's vibrational state, and then the characteristic function $\chi(\alpha) = \langle\psi|\hat{D}(\alpha)|\psi\rangle$ [31] can be obtained by measuring the expectation value $\langle\hat{Z}\rangle$, and the specific expression is as follows,

$$\begin{aligned}\langle\hat{Z}\rangle &= \langle e^{i\theta}\hat{D}(-\alpha) + e^{-i\theta}\hat{D}(\alpha)\rangle/2 \\ &= \cos(\theta)\text{Re}[\chi(\alpha)] + \sin(\theta)\text{Im}[\chi(\alpha)],\end{aligned} \quad (15)$$

where $\theta$ is the rotation angle of the electron spin state on the Bloch sphere, and $\phi$ is the phase of the carrier. The imaginary part of the characteristic function, $\text{Im}[\chi(\alpha)]$, is measured using a carrier with $\theta=\pi/2$, while the real part, $\text{Re}[\chi(\alpha)]$, is measured using a carrier with $\theta=0$. The specific implementation process can be referred to the reports of Flühmann and Home[31].

By measuring the characteristic function of the quantum state of the ion's motion and performing a two-dimensional Fourier transform on it, the Wigner function of the quantum phonon laser state can be reconstructed, thus realizing a complete characterization of the quantum state. The Wigner function not only reveals the coherence and phase information of the quantum state but also reflects its nonclassical properties, such as squeezing and entanglement.

## 5. Summary

The single-ion phonon laser state in the quantum regime is theoretically simulated based on a three-level atomic model. The steady-state phase diagram of the phonon laser state is obtained by solving the master equation. The Wigner quasiprobability distribution and second-order correlation properties of the phonon laser state are analyzed. In addition, although the phase diagrams of the three-level model and the two-level model cannot be compared under the same parameters, a significant deviation is observed when comparing their respective phonon laser generation thresholds. This indicates that the threshold conditions are model-dependent and cannot be simply interchanged.

By optimizing the interaction strength between the blue-sideband laser and the ion, the controllable excitation of the coherent phonon state and the precise control of the vibrational quantum state can be realized, and the stable phonon laser output with Poisson statistical characteristics can be obtained. In the experimental scheme, the characteristic function is used to measure the quantum state of the phonon laser state, which can not only accurately extract the coherence and phase information of the phonon laser state, but also deeply analyze the distribution characteristics of the quantum state. These findings not only provide a new idea for the physical realization of quantum phonon laser states, but also open up a new way for phonon state manipulation in quantum information processing and precision measurement technology based on trapped ions.